\documentclass[12pt]{article}

\usepackage{soul}

\usepackage{tikz}
\usepackage{epsfig,latexsym,amsfonts,amsmath,amsthm,amssymb,amsbsy,multirow,slashed,wasysym,textcomp,subfigure,wrapfig,datetime,comment,mathtools,cancel,cite,twistor,mathrsfs}
\usepackage[hidelinks]{hyperref}
\usepackage[font={footnotesize},bf]{caption}

\def\ee{\end{equation}}
\def\be{\begin{equation}}
\def\bea{\begin{eqnarray}}
\def\eea{\end{eqnarray}}
\newcommand{\beq}{\begin{eqnarray}}
\newcommand{\eqq}{\end{eqnarray}}
 \newcommand{\badat}{\begin{alignedat}}
 \newcommand{\eadat}{\end{alignedat}}

\newcommand{\eal}[1]{\be \begin{aligned} #1 \end{aligned}\end{equation}} 
\newcommand{\eqn}[1]{\be #1 \end{equation}} 
\newcommand{\eqa}[1]{\bea  #1\end{eqnarray}}

\renewcommand{\d}{\mathrm{d}}

\newcommand{\hphi}{\hat{\phi}}
\newcommand{\hpsi}{\hat{\psi}}
\long\def\new#1\endnew{{\bf #1}}		
\long\def\del#1\enddel{}
\def\eps{\epsilon }
\def\del{\partial}

\def\re{\mathrm{e}}

\def\stp{{\Sigma_{\tau+}}}
\def\stm{{\Sigma_{\tau-}}}
\def\st{{\Sigma_\tau}}
\def\s{\sigma }

\def\l{\lambda }
\def \hp{ \hat \Phi}

\def\ct{{\mathcal{CT}^2}}
\def\adz{AdS$_3/\mathbb{Z}$}

\usepackage{color}

\newcommand{\pink}[1]{\textcolor{\pink}{#1}}

\definecolor{dblue}{rgb}{0.2,0.50,0.80}

\def\S{\mathcal{S}}

\def\D{{\Delta}}

\def\G{{\Gamma}}

\def\K{\mathbb{K}}
\def\AdS{\text{AdS}}

\def\C{{\cal C}}

\def\s{ {\sigma} }
\def\T{{\cal T}}

\def\scrj{{\mathcal{I}}}
\def\scrj{{\mathcal{J}}}
\def\d{\delta}
\def\hp{{\bf \Phi}}
 \def\e{\epsilon}
 
 \newcommand{\csch}{\mathrm{csch}\hspace{.02in}}
\def\slw{$\SL(2,{\mathbb R})_L \times \SL(2,{\mathbb R})_R $}
\def\hs{{\hat \s}}
\begin{document}
\begin{titlepage}
\unitlength = 1mm~\\
\vskip 3cm
\begin{center}

{\LARGE{ No-Boundary State for  Klein Space }}

\vspace{0.8cm}
Walker Melton, Atul Sharma, Andrew Strominger and Tianli Wang \\
\vspace{1cm}

{\it  Center for the Fundamental Laws of Nature,\\ Harvard University, Cambridge, MA, USA} \\

\vspace{0.8cm}

\begin{abstract}
Analytic continuation from $(3,1)$ signature Minkowski to $(2,2)$ signature Klein space has emerged as  a useful tool for the  understanding of scattering amplitudes and flat space holography.  Under this continuation, past and future null infinity merge into a single boundary ($\scrj$) which is the product of a  null line with a $(1,1)$ signature torus. The Minkowskian $\S$-matrix  continues to a Kleinian  $\S$-vector which in turn may be represented by a Poincar\'e-invariant vacuum state  $|\C\rangle$ in the Hilbert space built on $\scrj$.  $|\C \rangle$ contains all information about $\S$ in a novel,  repackaged form. We give an explicit construction of $|\C\rangle$ in a Lorentz/conformal basis for a free massless scalar. $\scrj$ separates into two halves $\scrj_\pm $ which are the asymptotic null boundaries of the regions timelike and spacelike separated from the origin. $|\C\rangle$ is shown to be  a maximally entangled state in the product of the $\scrj_\pm $ Hilbert spaces.
 \end{abstract}

\end{center}

\end{titlepage}

\tableofcontents

\section{Introduction}

Analytic continuation to Euclidean space has been a remarkably effective  tool for analyzing both perturbative and nonperturbative quantum field theory  for the better part of the last  century \cite{Schwinger:1958mma}. In recent decades,  remarkable  insights have been obtained by alternate analytic continuations to complex split-signature spacetimes\cite{Travaglini:2022uwo, Heckman:2022peq,Britto:2005fq,Witten:2003nn}\footnote{The analytic continuations employed are often described as a complexification of momenta, but  can be equivalently geometrically viewed as a continuation to split signature.}.  This complimentary approach  is effective in part because  asymptotic scattering states can be both on-shell and  well-controlled by their analytic properties. Moreover it strongly resonates with twistor theory \cite{Penrose:1972ia,Mason:2005qu,ajm1180135553,Mason:2022hly,LeBrun:2005qf}
which also involves a complexification to split signature. 

Split signature naturally  arises in the celestial approach  to flat space holography.  
In this approach  flat space is foliated by hyperbolic slices, and one endeavors to uplift the well-understood holographic dictionary for hyperbolic spacetimes to a holographic dictionary for flat spacetimes \cite{deBoer:2003vf,Strominger:2017zoo,Ball:2019atb,Pasterski:2021raf,Melton:2023,Iacobacci:2022yjo}.
$(2,2)$ signature Klein space ($\K^{2,2}$) has an especially simple foliation by Lorentzian \adz\ slices and provides a natural starting point for this enterprise\cite{Atanasov:2021oyu, Melton:2023bjw,Melton:2024jyq,Melton:2024gyu}.   


A salient feature of Klein space is that null infinity ($\scrj$) has only a single connected component, which is the product of a null line with the celestial torus $\ct$.\footnote{The compactification  of $\scrj$ with temporal and spatial infinity $i_\pm$ added is an $S^3$.}  Since there are no separate in and out regions, we cannot have an $\S$-matrix in the ordinary sense \cite{Atanasov:2021oyu}. Instead we have an $\S$-vector, which may be viewed as quantum state in the Hilbert space on $\scrj$ and will be denoted $|\C\rangle$. This  state is a Kleinian analog of  the Hartle-Hawking no-boundary de Sitter state \cite{Hartle:1983ai}. It  contains all the data of the Minkowski $\S$-matrix, but analytically continued and repackaged in a novel and illuminating  manner. 

In this paper, we construct the state $|\C\rangle$ exactly for a free massless scalar. Under analytic continuation from Minkowski to Klein space, the Lorentz group $\SO(3, 1)$ becomes  \slw\ (which acts as the conformal group on $\ct$) \cite{Atanasov:2021oyu,Melton:2023hiq}.    Its orbits are  \adz\ slices ending  on the torus $\T_0$  at the intersection of the light cone of the origin with  $\scrj$. We construct a  covariant symplectic form on the classical phase space of the scalar field as \adz\ integrals of the conserved bilinear symplectic current.  $\T_0$ divides $\scrj$ into two halves, denoted $\scrj_-$ and $\scrj_+$, which are the two asymptotic limits of the \adz\ slices.  This symplectic form provides a basis for the construction and quantization of the $\scrj_\pm$ Hilbert spaces. 

The $\scrj_\pm$ Hilbert spaces are built on conformally invariant vacua and  manifestly decompose into representations of the conformal group.  There is no translation invariant state in either of these spaces.  However the Hilbert space on all of null infinity $\scrj=\scrj_-\cup \scrj_+$ is the  tensor product of those on $\scrj_\pm$.  We construct a fully Poincar\'e-invariant `vacuum' state $|\C\rangle$\footnote{We do not prove uniqueness.} in this  space, and show that it maximally entangles the $\scrj_\pm$ Hilbert spaces.  Our construction  resembles in some aspects  those of the Minkowski vacuum in the tensor product of the left and right Rindler Hilbert spaces, or the de Sitter vacuum in the north and south static patch Hilbert spaces \cite{Cotler:2023xku}.\footnote{As well as the conformally covariant construction of 
the Minkowski vacuum in \cite{cmsw}.}

A massless  particle incoming from $\scrj_-$ will reflect through the origin of Klein space and exit at $\scrj_+$. Similarly incoming waves on $\scrj_-$ will be reflected  and exit at $\scrj_+$. Here we construct the `clock matrix' $\C$  which evolves quantum states from  $\scrj_-$ to  $\scrj_+$ and is linearly related to $|\C \rangle$.
$\C$ and the $\S$-matrix are also linearly related. However the natural bases for these two objects are quite different and the relation is interestingly non-trivial \cite{Crawley:2021ivb}. It requires  the map between  tensor product states on the Kleinian $\scrj_+\cup\scrj_-$ and the Minkowskian $\scri^+\cup\scri^-$, to  which we hope to return to in the future.  

 Several decades ago, de Boer and Solodhukin \cite{deBoer:2003vf} hypothesized that a holographic dictionary for 4D flat space might  be derived by 
foliating it with AdS$_3$/dS$_3$ slices and uplifting the  hyperbolic holographic  dictionary  to flat space. Over the 
intervening decades, exploiting a variety of new inputs, a considerable number of entries in this  flat holography dictionary have been derived. This paper is motivated by and is a step forward in this broader program. 

We begin in Section 2 with preliminaries and notation. Section 3 constructs the classical symplectic form. Quantization on $\scrj_\pm$ is in section 4. Section 5 constructs the Poincar\'e invariant vacuum $|\C\rangle$. Section 6 discusses the relations between the Minkowski $\S$-matrix, its representation as a state $|\S\rangle $ in the tensor product of the $\scri^\pm$ Hilbert spaces, the $\C$-matrix  and $|\C\rangle$.

\begin{figure}[h!]
\begin{center}
\tikzset{every picture/.style={line width=0.75pt}} 

\begin{tikzpicture}[x=0.75pt,y=0.75pt,yscale=-1.25,xscale=1.25]

\draw  [draw opacity=0][fill={rgb, 255:red, 242; green, 242; blue, 242 }  ,fill opacity=1 ] (188.56,45.82) -- (377.44,235.04) -- (188.56,235.04) -- cycle ;
\draw [color={rgb, 255:red, 4; green, 103; blue, 220 }  ,draw opacity=1 ]   (188.29,112) .. controls (221.29,112.57) and (244.86,120.57) .. (284,141.43) ;
\draw [color={rgb, 255:red, 4; green, 103; blue, 220 }  ,draw opacity=1 ]   (188.29,152) .. controls (220.29,153) and (245.43,152.29) .. (284,141.43) ;
\draw [color={rgb, 255:red, 4; green, 103; blue, 220 }  ,draw opacity=1 ]   (188.43,85.29) .. controls (238.43,88.29) and (251.46,115.5) .. (284.46,142.21) ;
\draw [color={rgb, 255:red, 4; green, 103; blue, 220 }  ,draw opacity=1 ]   (188.29,198) .. controls (223.71,197.57) and (255.43,168.29) .. (284,141.43) ;
\draw [color={rgb, 255:red, 255; green, 0; blue, 113 }  ,draw opacity=1 ]   (311.29,235) .. controls (310.74,202) and (304.96,180.39) .. (284.08,141.23) ;
\draw [color={rgb, 255:red, 255; green, 0; blue, 113 }  ,draw opacity=1 ]   (273.29,235) .. controls (272.31,203) and (273.16,179.79) .. (284.08,141.23) ;
\draw [color={rgb, 255:red, 255; green, 0; blue, 113 }  ,draw opacity=1 ]   (339.29,235) .. controls (336.32,185) and (310.05,173.79) .. (283.29,140.77) ;
\draw [color={rgb, 255:red, 255; green, 0; blue, 113 }  ,draw opacity=1 ]   (226.29,235) .. controls (226.75,199.57) and (257.12,169.78) .. (284.08,141.23) ;
\draw    (284,141.43) -- (188.56,235.04) ;
\draw [color={rgb, 255:red, 128; green, 128; blue, 128 }  ,draw opacity=1 ][fill={rgb, 255:red, 128; green, 128; blue, 128 }  ,fill opacity=1 ]   (189.46,44.21) -- (211.29,22.39) ;
\draw [color={rgb, 255:red, 128; green, 128; blue, 128 }  ,draw opacity=1 ][fill={rgb, 255:red, 128; green, 128; blue, 128 }  ,fill opacity=1 ]   (285.08,140.23) -- (291.31,134) ;
\draw [color={rgb, 255:red, 128; green, 128; blue, 128 }  ,draw opacity=1 ][fill={rgb, 255:red, 128; green, 128; blue, 128 }  ,fill opacity=1 ]   (195.12,42.55) -- (229.71,77.14) ;
\draw [shift={(193,40.43)}, rotate = 45] [fill={rgb, 255:red, 128; green, 128; blue, 128 }  ,fill opacity=1 ][line width=0.08]  [draw opacity=0] (7.14,-3.43) -- (0,0) -- (7.14,3.43) -- (4.74,0) -- cycle    ;
\draw   (188.56,45.82) -- (377.44,235.04) -- (188.56,235.04) -- cycle ;
\draw [color={rgb, 255:red, 245; green, 166; blue, 35 }  ,draw opacity=1 ]   (188.29,141.71) .. controls (190.64,141.71) and (191.82,142.89) .. (191.82,145.25) .. controls (191.82,147.61) and (193,148.79) .. (195.36,148.79) .. controls (197.71,148.79) and (198.89,149.97) .. (198.89,152.32) .. controls (198.89,154.68) and (200.07,155.86) .. (202.43,155.86) .. controls (204.78,155.86) and (205.96,157.04) .. (205.96,159.39) .. controls (205.96,161.75) and (207.14,162.93) .. (209.5,162.93) .. controls (211.85,162.93) and (213.03,164.11) .. (213.03,166.46) .. controls (213.03,168.82) and (214.21,170) .. (216.57,170) .. controls (218.93,169.99) and (220.11,171.17) .. (220.11,173.53) .. controls (220.1,175.89) and (221.28,177.07) .. (223.64,177.07) .. controls (226,177.07) and (227.18,178.25) .. (227.18,180.61) .. controls (227.18,182.96) and (228.36,184.14) .. (230.71,184.14) .. controls (233.07,184.14) and (234.25,185.32) .. (234.25,187.68) .. controls (234.25,190.03) and (235.43,191.21) .. (237.78,191.21) .. controls (240.14,191.21) and (241.32,192.39) .. (241.32,194.75) .. controls (241.32,197.1) and (242.5,198.28) .. (244.85,198.28) .. controls (247.21,198.28) and (248.39,199.46) .. (248.39,201.82) .. controls (248.39,204.18) and (249.57,205.36) .. (251.93,205.35) .. controls (254.29,205.35) and (255.47,206.53) .. (255.46,208.89) .. controls (255.46,211.25) and (256.64,212.43) .. (259,212.42) .. controls (261.36,212.42) and (262.54,213.6) .. (262.53,215.96) .. controls (262.53,218.32) and (263.71,219.5) .. (266.07,219.5) .. controls (268.42,219.5) and (269.6,220.68) .. (269.6,223.03) .. controls (269.6,225.39) and (270.78,226.57) .. (273.14,226.57) .. controls (275.49,226.57) and (276.67,227.75) .. (276.67,230.1) .. controls (276.67,232.46) and (277.85,233.64) .. (280.21,233.64) -- (282,235.43) -- (282,235.43) ;
\draw [shift={(231.4,184.82)}, rotate = 45] [fill={rgb, 255:red, 245; green, 166; blue, 35 }  ,fill opacity=1 ][line width=0.08]  [draw opacity=0] (8.93,-4.29) -- (0,0) -- (8.93,4.29) -- (5.93,0) -- cycle    ;
\draw [color={rgb, 255:red, 245; green, 166; blue, 35 }  ,draw opacity=1 ]   (282,235.43) .. controls (282,233.07) and (283.18,231.89) .. (285.54,231.89) .. controls (287.89,231.89) and (289.07,230.71) .. (289.07,228.36) .. controls (289.07,226) and (290.25,224.82) .. (292.61,224.82) .. controls (294.96,224.82) and (296.14,223.64) .. (296.14,221.29) .. controls (296.14,218.93) and (297.32,217.75) .. (299.68,217.75) .. controls (302.03,217.75) and (303.21,216.57) .. (303.21,214.22) .. controls (303.21,211.86) and (304.39,210.68) .. (306.75,210.68) .. controls (309.11,210.68) and (310.29,209.5) .. (310.28,207.14) .. controls (310.28,204.78) and (311.46,203.6) .. (313.82,203.61) .. controls (316.18,203.61) and (317.36,202.43) .. (317.36,200.07) .. controls (317.36,197.72) and (318.54,196.54) .. (320.89,196.54) .. controls (323.25,196.54) and (324.43,195.36) .. (324.43,193) .. controls (324.43,190.65) and (325.61,189.47) .. (327.96,189.47) -- (330.29,187.14) -- (330.29,187.14) ;
\draw [shift={(302.4,215.03)}, rotate = 315] [fill={rgb, 255:red, 245; green, 166; blue, 35 }  ,fill opacity=1 ][line width=0.08]  [draw opacity=0] (8.93,-4.29) -- (0,0) -- (8.93,4.29) -- (5.93,0) -- cycle    ;
\draw [color={rgb, 255:red, 245; green, 166; blue, 35 }  ,draw opacity=1 ]   (188.29,141.71) .. controls (188.29,139.36) and (189.47,138.18) .. (191.82,138.18) .. controls (194.18,138.18) and (195.36,137) .. (195.36,134.64) .. controls (195.36,132.29) and (196.54,131.11) .. (198.89,131.11) .. controls (201.25,131.11) and (202.43,129.93) .. (202.43,127.57) .. controls (202.43,125.22) and (203.61,124.04) .. (205.96,124.04) .. controls (208.32,124.04) and (209.5,122.86) .. (209.5,120.5) .. controls (209.5,118.15) and (210.68,116.97) .. (213.03,116.97) .. controls (215.39,116.97) and (216.57,115.79) .. (216.57,113.43) .. controls (216.57,111.07) and (217.75,109.89) .. (220.11,109.89) .. controls (222.46,109.89) and (223.64,108.71) .. (223.64,106.36) .. controls (223.64,104) and (224.82,102.82) .. (227.18,102.82) .. controls (229.53,102.82) and (230.71,101.64) .. (230.71,99.29) .. controls (230.71,96.93) and (231.89,95.75) .. (234.25,95.75) -- (236.57,93.43) -- (236.57,93.43) ;
\draw [shift={(215.12,114.88)}, rotate = 135] [fill={rgb, 255:red, 245; green, 166; blue, 35 }  ,fill opacity=1 ][line width=0.08]  [draw opacity=0] (8.93,-4.29) -- (0,0) -- (8.93,4.29) -- (5.93,0) -- cycle    ;
\draw [color={rgb, 255:red, 128; green, 128; blue, 128 }  ,draw opacity=1 ][fill={rgb, 255:red, 128; green, 128; blue, 128 }  ,fill opacity=1 ]   (379.46,233.21) -- (400.82,211.86) ;
\draw [color={rgb, 255:red, 128; green, 128; blue, 128 }  ,draw opacity=1 ][fill={rgb, 255:red, 128; green, 128; blue, 128 }  ,fill opacity=1 ]   (290.43,139.12) -- (325.02,173.71) ;
\draw [shift={(288.31,137)}, rotate = 45] [fill={rgb, 255:red, 128; green, 128; blue, 128 }  ,fill opacity=1 ][line width=0.08]  [draw opacity=0] (7.14,-3.43) -- (0,0) -- (7.14,3.43) -- (4.74,0) -- cycle    ;
\draw [color={rgb, 255:red, 128; green, 128; blue, 128 }  ,draw opacity=1 ][fill={rgb, 255:red, 128; green, 128; blue, 128 }  ,fill opacity=1 ]   (286.16,134.59) -- (245.71,94.14) ;
\draw [shift={(288.29,136.71)}, rotate = 225] [fill={rgb, 255:red, 128; green, 128; blue, 128 }  ,fill opacity=1 ][line width=0.08]  [draw opacity=0] (7.14,-3.43) -- (0,0) -- (7.14,3.43) -- (4.74,0) -- cycle    ;
\draw [color={rgb, 255:red, 128; green, 128; blue, 128 }  ,draw opacity=1 ][fill={rgb, 255:red, 128; green, 128; blue, 128 }  ,fill opacity=1 ]   (381.16,227.59) -- (340.71,187.14) ;
\draw [shift={(383.29,229.71)}, rotate = 225] [fill={rgb, 255:red, 128; green, 128; blue, 128 }  ,fill opacity=1 ][line width=0.08]  [draw opacity=0] (7.14,-3.43) -- (0,0) -- (7.14,3.43) -- (4.74,0) -- cycle    ;
\draw [color={rgb, 255:red, 128; green, 128; blue, 128 }  ,draw opacity=1 ][fill={rgb, 255:red, 128; green, 128; blue, 128 }  ,fill opacity=1 ]   (208.43,28.12) -- (277.17,96.86) ;
\draw [shift={(206.31,26)}, rotate = 45] [fill={rgb, 255:red, 128; green, 128; blue, 128 }  ,fill opacity=1 ][line width=0.08]  [draw opacity=0] (7.14,-3.43) -- (0,0) -- (7.14,3.43) -- (4.74,0) -- cycle    ;
\draw [color={rgb, 255:red, 128; green, 128; blue, 128 }  ,draw opacity=1 ][fill={rgb, 255:red, 128; green, 128; blue, 128 }  ,fill opacity=1 ]   (394.16,213.86) -- (293.17,112.86) ;
\draw [shift={(396.29,215.98)}, rotate = 225] [fill={rgb, 255:red, 128; green, 128; blue, 128 }  ,fill opacity=1 ][line width=0.08]  [draw opacity=0] (7.14,-3.43) -- (0,0) -- (7.14,3.43) -- (4.74,0) -- cycle    ;
\draw  [draw opacity=0] (317.12,174.42) .. controls (317.06,174.48) and (317.01,174.54) .. (316.95,174.59) .. controls (298.83,192.53) and (269.6,192.39) .. (251.65,174.27) .. controls (233.81,156.25) and (233.85,127.23) .. (251.68,109.27) -- (284.46,141.78) -- cycle ; \draw [color={rgb, 255:red, 0; green, 0; blue, 0 }  ,draw opacity=1 ]   (317.12,174.42) .. controls (317.06,174.48) and (317.01,174.54) .. (316.95,174.59) .. controls (298.83,192.53) and (269.6,192.39) .. (251.65,174.27) .. controls (234.43,156.88) and (233.87,129.25) .. (249.88,111.19) ; \draw [shift={(251.68,109.27)}, rotate = 128.32] [fill={rgb, 255:red, 0; green, 0; blue, 0 }  ,fill opacity=1 ][line width=0.08]  [draw opacity=0] (10.72,-5.15) -- (0,0) -- (10.72,5.15) -- (7.12,0) -- cycle    ; 
\draw  [fill={rgb, 255:red, 0; green, 0; blue, 0 }  ,fill opacity=1 ] (281.76,141.34) .. controls (281.76,140.15) and (282.72,139.18) .. (283.92,139.18) .. controls (285.11,139.18) and (286.08,140.15) .. (286.08,141.34) .. controls (286.08,142.54) and (285.11,143.51) .. (283.92,143.51) .. controls (282.72,143.51) and (281.76,142.54) .. (281.76,141.34) -- cycle ;

\draw (279,96.83) node [anchor=north west][inner sep=0.75pt]  [color={rgb, 255:red, 0; green, 0; blue, 0 }  ,opacity=1 ]  {$\mathcal{J}$};
\draw (233,73.83) node [anchor=north west][inner sep=0.75pt]  [color={rgb, 255:red, 0; green, 0; blue, 0 }  ,opacity=1 ]  {$\mathcal{J}_{+}$};
\draw (327,168.83) node [anchor=north west][inner sep=0.75pt]  [color={rgb, 255:red, 0; green, 0; blue, 0 }  ,opacity=1 ]  {$\mathcal{J}_{-}$};
\draw (171,29.83) node [anchor=north west][inner sep=0.75pt]  [color={rgb, 255:red, 0; green, 0; blue, 0 }  ,opacity=1 ]  {$i_{+}$};
\draw (381.46,230.61) node [anchor=north west][inner sep=0.75pt]  [color={rgb, 255:red, 0; green, 0; blue, 0 }  ,opacity=1 ]  {$i_{-}$};
\draw (244.29,239) node [anchor=north west][inner sep=0.75pt]  [font=\scriptsize] [align=left] {temporal origin};
\draw (172.43,175.02) node [anchor=north west][inner sep=0.75pt]  [font=\scriptsize,rotate=-269.73] [align=left] {spatial origin};
\draw (289,123.83) node [anchor=north west][inner sep=0.75pt]  [font=\footnotesize,color={rgb, 255:red, 0; green, 0; blue, 0 }  ,opacity=1 ]  {$\mathcal{T}_{0}$};
\draw (223,127.4) node [anchor=north west][inner sep=0.75pt]  [color={rgb, 255:red, 0; green, 0; blue, 0 }  ,opacity=1 ]  {$\mathcal{C}$};

\end{tikzpicture}
	\end{center}
	\caption{ Toric Penrose diagram for signature $(2,2)$ Klein space. 45$^{\rm o}$ lines are null as usual. The red (blue) lines are \adz\ hypersurfaces of constant positive (negative) $X^2$.  A Lorentzian torus is fibered over every point in the diagram. The spacelike cycle of the torus degenerates along the vertical spatial origin, while the timelike cycle degenerates along the horizontal temporal origin. Neither cycle degenerates at null infinity $\scrj$ \cite{Atanasov:2021oyu}. The intersection of $\scrj$ with the light cone of the origin is the  torus $\T_0$ dividing  $\scrj$ into the two halves $\scrj_\pm$.  The wavy orange line depicts a light ray entering from $\scrj_-$,  reflecting through the temporal and spatial origins and emerging on $\scrj_+$. Field modes specified with incoming data on $\scrj_-$ are similarly doubly reflected up to $\scrj_+$. The clock matrix $\C$ evolves quantum states from $\scrj_-$ to $\scrj_+$ around $\T_0$.} \label{fig:1}\end{figure}
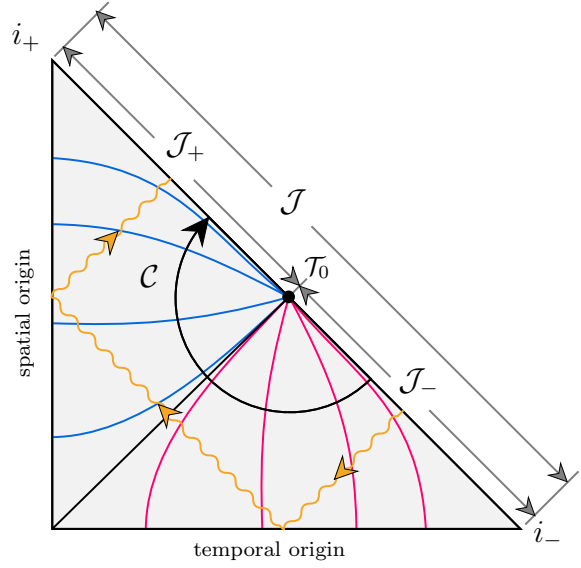

\section{Review of Klein space}

In this section we recall a  few salient features of Klein space and massless scalar conformal primary wavefunctions  \cite{Melton:2024jyq}.

The flat metric in Cartesian coordinates on  $\K^{2,2}$ is 
\be \rd s^2=-(\rd X^0)^2-(\rd X^1)^2+(\rd X^2)^2+(\rd X^3)^2\,.\ee
The geometry has a `$-$' spacelike wedge with $X^2>0$ and a `$+$' timelike wedge with $X^2<0$, each of which is foliated with a family \adz\ slices $\Sigma_{\tau-}$ and $\Sigma_{\tau+}$, as in Figure~\ref{fig:1}. 
In each wedge, the flat metric may be written 
 in hyperbolic coordinates as
\be \rd s_\pm^2=\mp \rd\tau^2\pm \tau^2 \rd s_{3}^2\,,\ee
where
\be \rd s_{3}^2=-\cosh^2\!\rho\, \rd\psi^2+\sinh^2\!\rho\, \rd\phi^2 +\rd\rho^2\,.\ee
This is the standard unit metric on \adz, where the $\mathbb Z$ acts by 
$\psi\mapsto \psi+2\pi$.  
In the spacelike  wedge we have  
\be X^0+iX^1=\tau\, \re^{i\phi}\sinh \rho\,,~~~~X^2+iX^3=\tau\, \re^{i\psi}\cosh \rho\,,\ee
with a similar formula  in the timelike wedge \cite{Melton:2023bjw}.

Defining the null vectors 
 \be \hat p^0+i\hat p^1=\re^{i\hat \phi},~~~~\hat p^2+i\hat p^3=\re^{i\hat \psi} \ee 
 where $\hat{\sigma}=(\hat\phi,\hat\psi)$ are coordinates on the Lorentzian celestial torus $\mathcal{CT}^2$, the scalar conformal primary wavefunctions are 
 \be\label{pdd}  \Phi^\D(\hat\s;X) =\frac{\G(\D)}{(\e-i\hat p\cdot X)^\D}\,.\ee
Here and below  we take the branch cut so that
\be \label{bcp}
(\eps-i x)^{-\Delta} = \re^{i\pi\Delta/2}|x|^{-\Delta}\Theta(x) + \re^{-i\pi\Delta/2}|x|^{-\Delta}\Theta(-x)
\ee
for real nonzero $x$.


\section{Symplectic product}

In this section, as a prelude to quantization,  we construct a symplectic form on the phase space of solutions of the free scalar wave equation. 

A symplectic form may be defined  from integrals of the one-form 
\be J^{12}=\Phi^1\overleftrightarrow {\p_\mu}\Phi^2 \rd X^\mu\,,\ee
which  obeys $\rd*J^{12}=0$ for any pair of solutions $\Phi^{1,2}$ of the free wave equation $\square \Phi=0$.  Given a surface $\Sigma$, a conserved\footnote{Of course, when interactions are included $J^{12}$ is typically at best conserved asymptotically.} symplectic product can be defined by
\be (\Phi^1|\Phi^2)_\Sigma=\int _\Sigma *J^{12}\,.\ee
A naive first guess might be to  choose $\Sigma$ to be $\scrj$ in Klein space. However, this will vanish because $\scrj$ is contractible (after adding the points $i_\pm$).  

Instead, we choose $\Sigma=\scrj_-$, which is half of $\scrj$. Specifying initial data on $\scrj_-$ determines a  unique solution of the free wave equation on all of $\K^{2,2}$ because the propagating wave reflects through the interior back up to $\scrj_+$, as in Figure~\ref{fig:1}. Hence the symplectic product so defined is a function on all of phase space. 
In free field theory, we  may equivalently   choose 
$\Sigma=\st$ to be any hyperbolic \adz\ surface of constant $\tau$ which ends on the torus $\mathcal{T}_0$. 
Since different choices of $\tau$ are homotopic, $(\Phi^1|\Phi^2)_\st$ is independent of $\tau$.  In practice it is useful to take $\Sigma=\stm$ to be a hyperbolic surface of large positive $X^2=\tau^2$ near $\scrj_-$ where interactions are suppressed. 

We now evaluate $(\Phi^1|\Phi^2)_\stm$ for the conformal primary wavefunctions. Restricted to the spacelike wedge, they are 
\be \label{spw} \Phi^\D|_{X^2>0}={\G(\D) \over \tau^\D}G_\D(\hs; \rho,\phi,\psi)\,, \ee
where
\be G_\D(\hs; \rho,\phi,\psi)=  \big(\e-i\cos(\hat \psi-\psi)\cosh \rho +i\cos(\hat \phi-\phi)\sinh \rho \big )^{-\D}\ee
are the weight $\D$ bulk-to-boundary propagators. For the rest of this work, we take $\D = 1+i\lambda$ for $\lambda \in \mathbb{R}$. 

These obey the convolution formula\footnote{Here $x$ and $\mathrm{D}^3x$ are abbreviations for $(\rho,\phi,\psi)$ and the associated measure.}
\begin{align}
        &\int_{\AdS_3/\Z} \mathrm{D}^3x\;G_{\D_1}(\hs_1;x) \, G_{\D_2}(\hs_2;x)\nonumber \\
&= \frac{2\pi^2e^{\pi\lambda_1}}{\lambda_1}\,\delta(\lambda_{12}) 
         \left(\frac{1}{\big(\sin\frac{\hphi_{12}+\hpsi_{12}}{2}\sin\frac{\hphi_{12}-\hpsi_{12}}{2}+i\eps\big)^{\Delta_1}}-\frac{1}{\big(\sin\frac{\hphi_{12}+\hpsi_{12}}{2}\sin\frac{\hphi_{12}-\hpsi_{12}}{2}-i\eps\big)^{\Delta_1}}\right) \nonumber\\
        &+ \frac{16\pi^3}{\lambda_1^2}\,\delta(\lambda_1+\lambda_2)\left[\delta^2(\hs_{12}) +\cosh\pi\lambda_1\delta^2(\hs_{12}+\pi)\right]\,.
    \end{align}
This is analogous to  similar formulae in Euclidean AdS \cite{Costa:2014kfa} and to the leaf two-point function in \cite{Melton:2024jyq}. Here $\hs_{ij}=\hs_i-\hs_j$, $\Delta_{ij}=\Delta_i-\Delta_j$, etc., and $\hs+\pi=(\hat\phi+\pi, \hat \psi+\pi)$ is the antipodal point to $\hs$. 
Using this, one finds the symplectic product in the spacelike wedge,
\begin{equation}
\begin{split}\label{spr}
     &(\Phi^{\D_1}|\Phi^{\D_2})_{\stm} = -\Delta_{12}\,\tau^{2-\D_1-\D_2}\Gamma(\Delta_1)\,\Gamma(\Delta_2)\int_{\stm} \mathrm{D}^3x\; G_{\D_1}(\hs_1 ;  x)\,G_{\D_2}(\hs_2 ; x)   \\
     &= -32\pi^4i\,\delta(\lambda_1+\lambda_2)\left[\coth\pi\lambda_1\delta^{(2)}(\hs_{12}+\pi) + \csch\pi\lambda_1\delta^{(2)}(\hs_{12})\right].
    \end{split}
\end{equation}
Here we take the unit normal on $\stm$ to be inward pointing which asymptotes to the future directed normal on $\scrj_-$.  

Going forward, it considerably simplifies the equations to employ a diagonal basis  which  transforms simply under the action of PT, which acts on $\ct$ by shifting $\hs\mapsto\hs+\pi$. This basis is 
\be 
\begin{split}
    \Phi^\l_E(\hs)&=N_E\left(\Phi^{1+i\l}(\hs)    +\Phi^{1+i\l}(\hs+\pi)\right)\,,\\
    \Phi^\l_O(\hs)&= N_O\left(\Phi^{1+i\l}(\hs)    -\Phi^{1+i\l}(\hs+\pi)\right)\,, 
\end{split}
\ee
where $E$ and $O$ stand for even and odd. The normalizations are given by
\be
N_E = \frac{1}{8\pi^2}\sqrt{\tanh\!\left(\frac{\pi|\lambda|}{2}\right)}\,,\qquad\ N_O = \frac{1}{8\pi^2}\sqrt{\coth\!\left(\frac{\pi|\lambda|}{2}\right)}\,.
\ee
These normalizations are chosen so that CPT acts by sending $\lambda \to -\lambda$. The symplectic products are 
\begin{equation}\label{EOsymp}
    \begin{split}
(\Phi_E^{\l_1}|\Phi_E^{\l_2})_{\stm}&=-i\hspace{.02in}\sgn(\lambda_1)\,\d(\l_1+\l_2)\left(\d^2(\hs_{12})+\d^2(\hs_{12}+\pi)\right)\,,\\
(\Phi_O^{\l_1}|\Phi_O^{\l_2})_{\stm}&=-i\hspace{.02in}\sgn(\lambda_1)\,\d(\l_1+\l_2)\left(\d^2(\hs_{12}+\pi)-\d^2(\hs_{12})\right)\,,\\
(\Phi_E^{\l_1}|\Phi_O^{\l_2})_{\stm} &=0\,.
\end{split}
\end{equation}

A similar construction can be used in the timelike wedge. Here  the normal to $\Sigma_{\tau_+}$ is taken to be outward pointing so that it also asymptotes to the future directed normal on $\scrj_+$. This gives an extra minus sign leading to 
\begin{equation}\label{fdg}
    \begin{split}
(\Phi_E^{\l_1}|\Phi_E^{\l_2})_{\stp}&=i\hspace{.02in}\sgn(\lambda_1)\,\d(\l_1+\l_2)\left(\d^2(\hs_{12})+\d^2(\hs_{12}+\pi)\right),\\
(\Phi_O^{\l_1}|\Phi_O^{\l_2})_{\stp}&=i\hspace{.02in}\sgn(\lambda_1)\,\d(\l_1+\l_2)\left(\d^2(\hs_{12}+\pi)-\d^2(\hs_{12})\right),\\
(\Phi_E^{\l_1}|\Phi_O^{\l_2})_{\stp} &=0\,.
\end{split}
\end{equation}
As a result, one confirms 
\be (\Phi^{\l_1}_{E,O}|\Phi^{\l_2}_{E,O})_{\scrj} =0 \ee
as expected from the contractibility of $\cJ$.

\section{Hilbert spaces on $\scrj_\pm$}
  
In this section we describe the quantum Hilbert spaces on the boundary regions $\scrj_\pm$, viewed as asymptotic limits of $\Sigma_{\tau_{\pm}}$. A priori, these spaces are independent. However they may be related by evolution through the bulk, much as incoming and outgoing states on $\scri^\pm$ in Minkowski space are related by the $\S$-matrix. 
  
Starting with  the spacelike wedge on the $\stm$ slices, we  promote the wave functions to modes of the field operator $\hp(\tau,x)$ via 
 \be \hp^\l_{E,O- }(\hs)=(\Phi^{\l}_{E,O}(\hs)|\hp)_{\stm} \ee where $\hs=(\hat \phi, \hat \psi)$. Multiplying \eqref{EOsymp} by $i$ these obey
 the quantum commutators 
 \be\label{-comm}
 \begin{split}[\Phi_{E-}^{\l_1},\Phi_{E-}^{\l_2}]_{\stm}&=\sgn(\lambda_1)\d(\l_1+\l_2)(\d^2(\hs_{12})+\d^2(\hs_{12}+\pi))\,,\cr
[\Phi_{O-}^{\l_1},\Phi_{O-}^{\l_2}]_{\stm}&=\sgn(\lambda_1)\d(\l_1+\l_2)(\d^2(\hs_{12}+\pi)-\d^2(\hs_{12}))\,,\cr
 [\Phi_{E-}^{\l_1},\Phi_{O-}^{\l_2}]_{\stm}&=0\,.~~~~~~~~~~~~~~~~~~~~~~~~~~~~~~~~~~~~~~~~~~
 \end{split}
 \ee
In a slight abuse of terminology,  the minus  Rindler  vacuum is defined by \be \hp_{E,O-}^{-\l} (\hs) |0_{R-}\rangle =0\qquad\forall\;\l > 0\,.\ee 
This vacuum is  manifestly invariant under \slw\ conformal transformations, but not under spacetime translations which shift $\l$ by $i$. 
 
 We define bulk hermitian conjugation of these operators  as usual by complex conjugation of the associated modes
 \be \hp^\dagger=\hp\,,~~~\hp^{\l}_{E,O-}(\hs)^\dagger=\hp^{-\l}_{E,O-}(\hs+\pi)\,,~~~~ |0_{R-}\rangle^\dagger = \langle 0_{R-}|\,,\ee
 where $\langle 0_{R-}|$ is defined by
 \be \langle 0_{R-}|\hp^{-\l}_{E,O-}(\hs)^\dagger =0\qquad\forall\;\lambda>0 \,.\ee
A  general multi-particle Fock state on $\scrj_-$ is
\be\label{cvv}            |\Psi_{J-}\rangle= \prod_{k=1}^m\frac{(\Phi_{E-}^{\lambda_k}(\hs_k))^{p_k}}{\sqrt{p_k!}}\prod_{\ell=m+1}^{m+n}\frac{(\Phi_{O-}^{\lambda_\ell}(\hs_\ell))^{p_\ell}}{\sqrt{p_\ell!}}|0_{R-}\rangle\,,\ee
where the $\l_k$ are all positive and $J$ is a multi-index.  These form a basis of  states supported  on $\scrj_-$ with inner product denoted\footnote{In order for $N_{IJ-}$ to be invertible we should restrict the $\hs_k$ to range over a single diamond of the celestial torus.}
\be  \langle \Psi_{I-} |\Psi_{J-}\rangle=N_{IJ-}\,.\ee

 A similar construction gives a Hilbert space on $\scrj_+$.  A corresponding  basis of operators is   \be \hp^\l_{E,O+}(\hs)=(\Phi_{E,O}^{\l}(\hs)|\hp)_{\stp}\,.  \ee 
  Here, as in \eqref{fdg}, we use the outward pointing normal in defining the symplectic product. These have commutators
\be\label{+comm}
\begin{split}
[\Phi_{E+}^{\l_1},\Phi_{E+}^{\l_2}]_{\stp}&=-\sgn(\lambda_1)\d(\l_1+\l_2)(\d^2(\hs_{12})+\d^2(\hs_{12}+\pi))\,,\cr
[\Phi_{O+}^{\l_1},\Phi_{O+}^{\l_2}]_{\stp}&=-\sgn(\lambda_1)\d(\l_1+\l_2)(\d^2(\hs_{12}+\pi)-\d^2(\hs_{12}))\,,\cr
[\Phi_{E+}^{\l_1},\Phi_{O+}^{\l_2}]_{\stm}&=0\,.~~~~~~~~~~~~~~~~~~~~~~~~~~~~~~~~~~~~~~~~~~
\end{split}
\ee
  The $+$ Rindler vacuum is then 
 \be \hp_{E,O+}^{\l} (\hs) |0_{R+}\rangle =0\qquad\forall\;\l > 0\,.\ee 
Note the opposite  sign choice for $\l$: this  follows the Rindler analogy where annihilation operators in the two Rindler wedges have opposite-signed boost eigenvalues.  A general $\scrj_+$ Fock state is
\be\label{cvvb}            |\Psi_{J+}\rangle= \prod_{k=1}^m\frac{(\Phi_{E+}^{\lambda_k}(\hs_k))^{p_k}}{\sqrt{p_k!}}\prod_{\ell=m+1}^{m+n}\frac{(\Phi_{O+}^{\lambda_\ell}(\hs_\ell))^{p_\ell}}{\sqrt{p_\ell!}}|0_{R+}\rangle\,,\ee
with inner product $\langle \Psi_{I+} |\Psi_{J+}\rangle=N_{IJ+}$.
 
It is interesting to construct a dilation generator $\mathcal{D}_-$ which generates dilations from the light cone of the origin at $\ln \tau=-\infty$ to $\scrj_-$ at $\ln \tau =\infty$. The expression is
\be \mathcal{D}_-={1 \over 2}\int_0^\infty \rd\l\int_\ct \rd^2\hs \;\l\,\big(\hp_{E-}^\l\hp_{E-}^{-\l} -\hp_{O-}^\l\hp_{O-}^{-\l}\big)\,.\ee
Using the commutator \eqref{-comm} one finds
\be [\mathcal{D}_-,\hp_{E,O-}^\l(\hs)]=-\lambda\hp_{E,O-}^\l(\hs)\,,\ee
implying the exponentiated form
\be \re^{-i\mathcal{D}_-\ln \tau}\hp_{E,O-}^\l(\hs)\re^{i\mathcal{D}_-\ln \tau}=\tau^{i\l} \hp_{E,O-}^\l(\hs)\,.\ee

\section{Poincar\'e invariant vacuum state $|\C\rangle$ on $\scrj$ }

Since $\scrj=\scrj_-\cup \scrj_+$,  a basis for the Hilbert space on $\scrj$ is given by sums of tensor products of $\scrj_-$ states of the form 
\eqref{cvv} with $\scrj_+$ states of the form \eqref{cvvb}. Among all these states is a special one that is Poincar\'e invariant, which is given by 
\be |\C \rangle= \re^{X}|0_R\rangle\,\ee
with 
\be |0_{R}\rangle = |0_{R+}\rangle |0_{R-}\rangle\,,\ee
\be \label{xep} X={1 \over 2}\int_0^\infty \rd\l\int_\ct \rd^2\hs \left(\hp_{E-}^\l\hp_{E+}^{-\l} - \hp_{O-}^\l\hp_{O+}^{-\l} \right)\,.\ee

Lorentz invariance of  $|\C \rangle$ follows immediately from Lorentz invariance of  
$X$  and $|0_R\rangle$.  To verify translation invariance we first note that 
\be 
(\hp^\l_{E,O+}+\hp^\l_{E,O-})|\C \rangle=0 \ee
with unrestricted real $\l$. The translation operator $P$ is linear and shifts $\l$ by $i$ in all operators. Since it does not mix up the two wedges it can be written in  the form \be P=f(\hp^\l_ {E,O-})-f(\hp^\l_{E,O+})\,,\ee
where $f$ is quadratic in the fields. The minus sign arises from the opposite signs in the $\hp^\l_{E,O+}$  vs  $\hp^\l_{E,O-}$
commutators in \eqref{-comm} and \eqref{+comm}. It follows immediately that 
\be f(\hp^\l_{E,O-})|\C \rangle=f(\hp^\l_{E,O}+)|\C \rangle\,,\ee
and \be P|\C \rangle=0\,.\ee
Hence $|\C \rangle$ is Poincar\'e invariant.

\section{Clock matrix }
In this section we discuss the relation of $|\C \rangle$ to the clock matrix which generates evolution from Kleinian $\scrj_-$ to $\scrj_+$, as well as to the Minkowski $\S$-matrix which generates evolution from Minkowskian $\scri^-$ to $\scri^+$. 

The Minkowski $\S$-matrix can be trivially encoded into a quantum state in the tensor product of the in and out Hilbert spaces. Let us choose a basis of incoming states $|A_I\rangle$ on $\scri^-$
obeying 
\be \langle A_I|A_J\rangle=\d_{IJ}\,,\ee
where $ \langle A_I|$ is the standard hermitian conjugate of $|A_I\rangle$. On $\scri^+$ we use an isomorphic basis obeying
\be \langle B_I|B_J\rangle=\d_{IJ}\,.\ee  The $\S$ matrix  in this basis is defined by
\be |B_I\rangle=\S_I^{~J}|A_J\rangle\,.\ee

As a way to repackage the scattering data, we can define a quantum state in the tensor product of the in and out Hilbert spaces using $\S_I^{~J}$,
\be |\S\rangle\vcentcolon= \d^{IK}\S_K^{~J}|B_I\rangle| A_J\rangle\,.\ee
To reproduce $\S_I^{~J}$ from  $|\S\rangle $, we simply take 
\be \delta^{KJ} \langle B_I |\langle A_K|\S\rangle = \S_I^{~J}\,.\ee

A similar construction pertains to Klein space. 
We may define a clock matrix by 
\be N^{KJ-} \langle \Psi_{I+} |\langle \Psi_{K-}|\C\rangle = \C_I^{~J}\,.\ee
In terms of $\C$ the Poincar\'e invariant vacuum state is  
\be |\C \rangle =N^{IK+}\C_K^{~J}|\Psi_{I+}\rangle| \Psi_{J-}\rangle\,. \ee
In our free field example we have simply 
\be \C\Phi^\l_{E,O-}\C^{-1}=-\Phi^\l_{E,O+}\,.\ee
Of course when interactions are included nonlinear corrections will appear. 

The states $|\S\rangle$ and $|\C\rangle$ maximally entangle the two  Hilbert spaces comprising  the tensor products from which they are constructed. Tracing over the out Hilbert space in Minkowski gives
\be \tr_B |\S\rangle\langle \S|=\S^{\dagger K}_L\S_K^{~I}\d^{LJ}|A_I\rangle\langle A_J|={\bf 1} \ee
for a unitary theory. Explicit computation for free field theory from \eqref{xep}  in Klein space gives 
\be \tr_+ |\C \rangle\langle \C |=\C^{\dagger K}_L\C_K^{~I}N^{LJ-}|\Psi_{I-}\rangle\langle \Psi_{J-}|={\bf1}\, .\ee
It is interesting to ask whether or not this maximal entanglement persists in  the presence of interactions. 

\section*{Acknowledgements}

The authors would like to thank Jordan Cotler, Matthew Dodelson, Matthew Heydeman, Lionel Mason, Noah Miller, Diandian Wang and Zixia Wei for useful conversations. This work was supported in part by NSF PHY-2207659, the Simons Collaboration on Celestial Holography, and the Gordon and Betty Moore Foundation and the John Templeton Foundation via the Black Hole Initiative.

\bibliographystyle{JHEP}
\bibliography{refs}


\end{document}